# Democratizing Artificial Intelligence in Healthcare: A Study of Model Development Across Two Institutions Incorporating Transfer Learning.

**Vikash Gupta**[1,†], **Holger Roth**[2], **Varun Buch**[3], **Marcio A.B.C. Rockenbach**[3], **Richard D White**[1], **Dong Yang**[2], **Olga Laur**[3], **Brian Ghoshhajra**[3], **Ittai Dayan**[3], **Daguang Xu**[2], **Mona G. Flores**[2], and **Barbaros Selnur Erdal**[1]

[1]Department of Radiology, Mayo Clinic, Jacksonville, Florida
[2]NVIDIA Inc, Santa Clara, California
[3]Department of Radiology, Massachusetts General Hospital, Boston, Massachusetts
[†]Corresponding Author: gupta.vikash@mayo.edu

## ABSTRACT

The training of deep learning models typically requires extensive data, which are not readily available as large well-curated medical-image datasets for development of artificial intelligence (AI) models applied in Radiology. Recognizing the potential for transfer learning (TL) to allow a fully trained model from one institution to be fine-tuned by another institution using a much small local dataset, this report describes the challenges, methodology, and benefits of TL within the context of developing an AI model for a basic use-case, segmentation of Left Ventricular Myocardium (LVM) on images from 4-dimensional coronary computed tomography angiography. Ultimately, our results from comparisons of LVM segmentation predicted by a model locally trained using random initialization, versus one training-enhanced by TL, showed that a use-case model initiated by TL can be developed with sparse labels with acceptable performance. This process reduces the time required to build a new model in the clinical environment at a different institution.

## 1 Introduction

The lack of large datasets of curated medical images severely limits the training of clinically useful deep neural networks. Compared to publicly available datasets in the computer vision community (e.g., ImageNet[1], Microsoft-COCO[2], Pascal-VOC[3]), accessible radiological datasets are relatively small. This is compounded by the need for relatively greater expertise in annotating radiological images, compared to general images. Consequently, many medical institutions will not have sufficient expert human resources to label the many medical images needed for training deep neural networks. As a result, transfer learning (TL) can play a significant role.

TL is a method where the weights of a previously trained model are used to initialize the otherwise random and inefficient training a new model as a form of knowledge transmission for a relatively similar task. As such, model training from the "pre-trained model"[4] can be accomplished on smaller datasets in order to improve model parameters for better performance.

Typically, if an institution intends to locally deploy a deep learning feature/workflow in its daily medical imaging practice, a radiologist needs to become dedicated to labeling a significant number of imaging datasets. Next, a team of deep learning scientists needs to preprocess the image annotations and design appropriate neural network architectures followed by ablation studies[5] for model verification. Unfortunately, this is a very time-consuming and resource-intensive process, inhibiting many institutions from engaging locally in artificial intelligence (AI) development. On the other hand, if a well-tested deep learning model from another institution is available, it can be adopted for use in model fine-tuning on a relatively smaller local dataset, thereby allowing institutions with fewer resources to also directly participate in AI development.

Although TL is a standard technique in deep learning associated with computer vision, it is not commonly used in medical imaging although researchers have successfully tried cross-domain TL (i.e., training models on datasets such as ImageNet and fine-tuning on radiological images)[6–10]. In computer vision, the application of TL is facilitated by the availability of multiple public datasets and pre-trained deep learning models (e.g., Inception-V3, ResNet, AlexNet). In addition, computer vision data are only two-dimensional, and image annotations are unambiguous, based on the appearances of everyday objects. On the other hand, despite general guidelines on annotating medical images, the quality of the labels is usually contingent on both radiologist-experience and image-quality levels. In addition, due to anatomic variability, two radiologists may disagree on annotation. This is crucial in TL as the quality of labels used for the initial training should correspond to the quality of labels relied on for fine-tuning. Thus, concordance between radiologists regarding annotating is essential. Unfortunately, across different healthcare institutions, the standards for imaging might still be different depending on the scanners, image reconstruction protocols, and the software used for local image viewing and/or labeling.

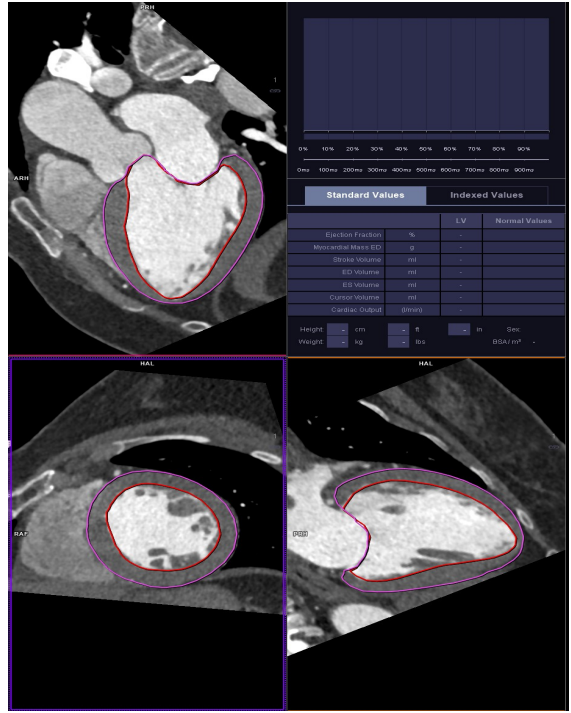

**Figure 1.** User interface (Sygno.via) where the annotations are made. The figure shows three conventional multiplanar views [top-left: 3-chamber; bottom-left: short-axis; bottom-right: 2-chamber]. The epicardial [purple] and endocardial [red] contours are initially generated automatically; the expert makes any needed adjustments on these contours for correction. The output table of volumetric results is also shown [upper right].

The goal of this report is to address these issues in TL as they might apply to a specific common clinical application: left ventricular (LV) segmentation. While they are not necessarily drawbacks or major hindrances to TL, their disregard may cause the advantages of TL to not be fully realized. We outline the different considerations prior to undertaking a TL project or designing TL workflow in medical imaging. To this end, we focus on a use-case related to the segmentation of four-dimensional (4D) (3 spatial dimensions and time dimension) dynamic cardiac computed tomography angiography (CCTA) images. In so doing, we also demonstrate that TL can be used to create a well-performing model by fine-tuning a neural network, rather than training one de novo, even when using a smaller number of images. Consequently, a model can either be matured more quickly or more fine-tuned through application of TL methodology.

## 2 Data

Institutional Review Board approval was obtained at both collaborating institutions, referred to as Site 1 (Massachusetts General Hospital) and Site 2 (Ohio State University College of Medicine). At Site 2, 95 cases of standard-of-care (between 01/01/2017 and 07/30/2019) 4D CCTA imaging (Somatom AS+ 128, Definition, or Force: Siemens Healthineers, Forchheim, Germany) were randomly selected. For each case, the retrospectively gated volumetric series was reconstructed at 20 uniformly distributed phases of the cardiac cycle, including end-diastole (ED) and end-systole (ES), using standard procedures[11,12]. For each series, the base transaxial slices contained 512 x 512 pixels with a spacing of 0.5 mm. The number of slices in each 4D CCTA image volume varied (146 to 249), but with a constant slice thickness of 0.7 mm.

Each 4D CCTA volume set was semi-automatically segmented using a commercial advanced image-processing software package

(syngo.via version vb30: Siemens Healthineers, Forchheim, Germany) running as a thin-client to a standard picture archiving and communication system (PACS). To derive LV Myocardium (LVM) segmentations, including LV cavity volumes, LV endocardial and epicardial surfaces were automatically delineated at both ED and ES, but corrected by an expert cardiovascular radiologist (RDW) for any errors prior to use [Figure 1].

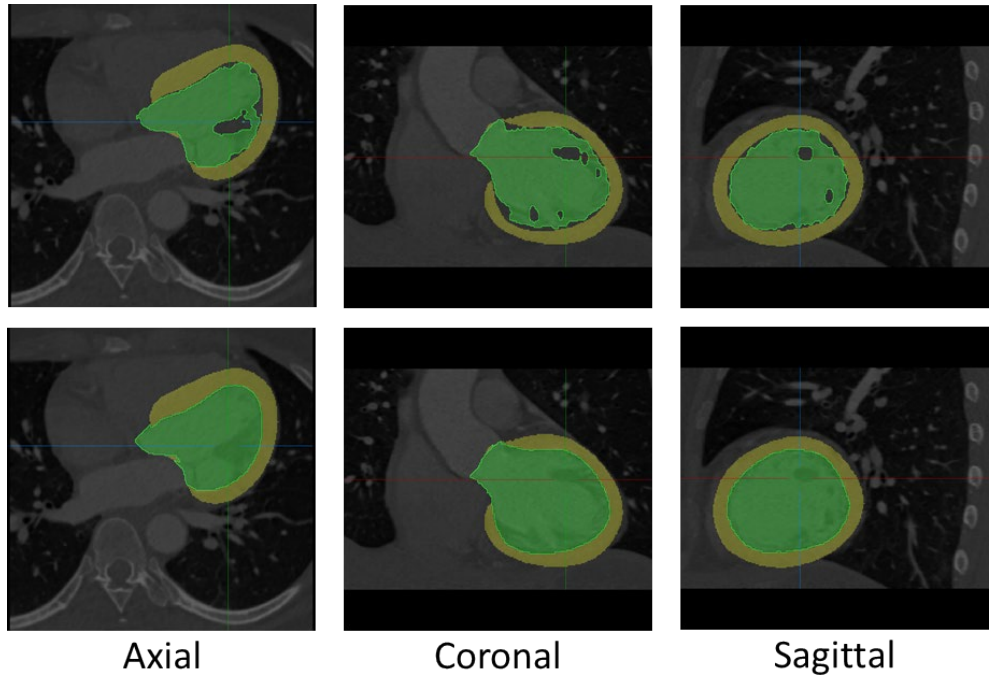

**Figure 2.** Holes in the LV blood-pool mask (green) and gaps between the blood pool and LVM mask (yellow) are initially present [**top row**]. These errors are rectified using the morphological image operations of dilation followed by erosion [**bottom row**].

## 3 Methods

### 3.1 Image Annotation

Generally, the clinically available tools are not well-suited for displaying and curating images for AI-based tasks[13]. One of the earliest TL-related challenges was confronted during the curation process and pertained to building an inter-site consensus about landmarks for image annotation. This was accomplished after forming important mutual agreements on : 1. treatment of papillary muscles (i.e., considered part of LV cavity); delineation of atrioventricular margin of LV cavity (i.e., 3D curve along mitral valve leaflets, including along anterior leaflet extending to posterior aortic root wall; and 3. inclusion of LV outflow tract (i.e., truncated with 3D curve along aortic valve leaflets) [Figure 1]. Consistency was confirmed in five cases from Site 2, which were blindly segmented by a Site 2 expert (RDW) separately from a Site 1 expert (OL), both applying the aforementioned segmentation guidelines.

In preparation for TL, we utilized a previously described 4D segmentation model architecture from collaborating Site 1[14]. The model was trained on comparable 4D CCTA images (Somatom Force and Flash: Siemens Healthineers, Forchheim, Germany) from 61 subjects. However, for each Site 1 case, local annotation was performed, in many cases, on every-other of the 20 phases of the cardiac cycle while including both ED and ES. On the other hand, Site 2 fine-tuned the Site 1 model on a sparse segmentation (i.e., only ED and ES phases labeled).

### 3.2 Preprocessing

In preprocessing, it is important to correct any errors (e.g., holes, gaps) in segmentation before using them for training the neural network. During the labeling process, the expert edits the automatically generated (by syngo.via) contours of the LVM; based on the edited contours, pixel-wise segmentation are produced using a flood-fill algorithm. However, as these segmentations are created automatically, holes in the cavity/blood-pool mask are typically present [Figure 2]. In addition, there is also commonly

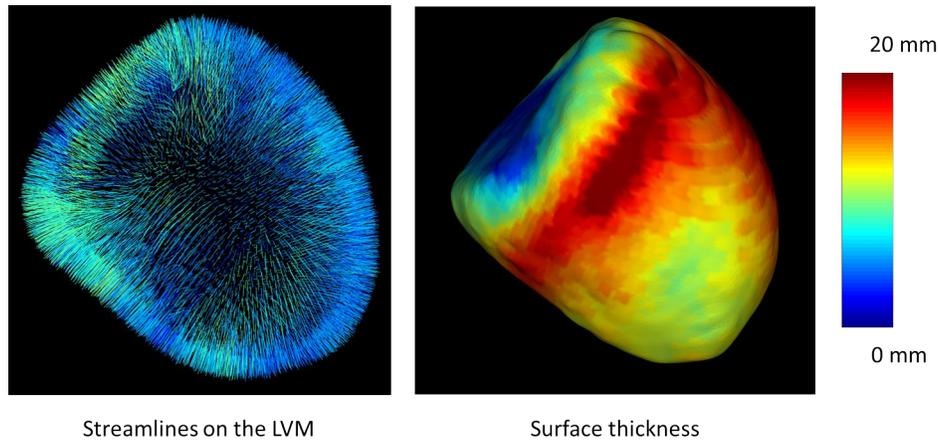

Streamlines on the LVM          Surface thickness

**Figure 3. Left**: The streamlines computed using differential equation 2. They start from the LVM outer surface and terminate at the interior surface. **Right**: Thickness measurements computed using calculations of the length of the streamlines (on the left) and overlaid on the outer surface of the LVM.

a gap between the blood-pool and the LVM mask. To correct for these defects, morphological operations (e.g., dilation followed by erosion) were used. While the dilation process closes internal holes, it causes unwanted expansion of the boundary of the mask. Thus, the dilation process is followed by an erosion of the external boundaries. The radius for dilation and erosion was chosen to be 5 pixels. These segmentation and subsequent corrections were done only for the ED and ES phases at Site 2. Once post-processing was completed, a segmentation image was created with labels 0 for background, 1 for blood-pool, and 2 for LVM. Using these labels, a 4D image-mask pair was created. To that end, all ED and ES images and corresponding masks were resampled to a 3D volume with 1-mm isometric spacing. These images consisted of 3D images acquired in 20 phases, one 3D image per phase. While the mask consisted of twenty 3D label arrays, other than for the ED and ES phases, the arrays for diastolic/filling and systolic/contraction phases were empty (zero arrays).

### 3.3 LVM Wall Thickness Computation
To evaluate the efficacy of the TL method, we use LVM wall thickness as a surrogate measure.

In order to compute the LVM wall thickness, we used a 3D parameterization technique[15]. The problem can be defined as analogous to a heat transfer problem. If there are two points in a connected body (genus-0 shape) which are at different fixed temperatures and the system is closed (no external heat), the system will reach a steady-state where the heat diffuses from a high temperature (source) to a low temperature (sink). During this process, a number of isothermal surfaces (heat gradients) will be established. At this point, isothermal lines can be computed by solving the following Laplace's equation:

$$\nabla^2 \Psi = \frac{\partial^2 \Psi}{\partial x^2} + \frac{\partial^2 \Psi}{\partial y^2} + \frac{\partial^2 \Psi}{\partial z^2}$$

, where $\Psi$ is the potential (or temperature) at any given point, and $x$, $y$, and $z$ are the dimensions. The heat flow lines are always orthogonal to the isothermal surfaces by construction and can be computed using the following differential equation:

$$\frac{\partial \Psi}{\partial t} = -\mu \nabla \Psi [X(t)]$$

, where $X$ is the location $(x, y, z)$, $t$ is time and $\mu$ is a normalizing parameter. The time t is not the temporal dimension in the cardiac phase cycle, but rather a time parameter in the heat flow equation which tracks the aforementioned flow of heat from the source to the sink. The length of these heat flow lines can be calculated, serving as representations of thickness of the LVM wall.

In order to compute the LVM wall thickness, the streamlines were computed by iteratively solving the differential equation 2 using the Jacobi technique[16]. Solving the differential equation gives the heat flow lines which can be used for calculating the thickness [Figure 3]. All streamlines start from the outer/epicardial surface of the LVM and terminate at the inner/endocardial surface. The thickness is calculated by adding the linear distances between the consecutive points on the streamline.

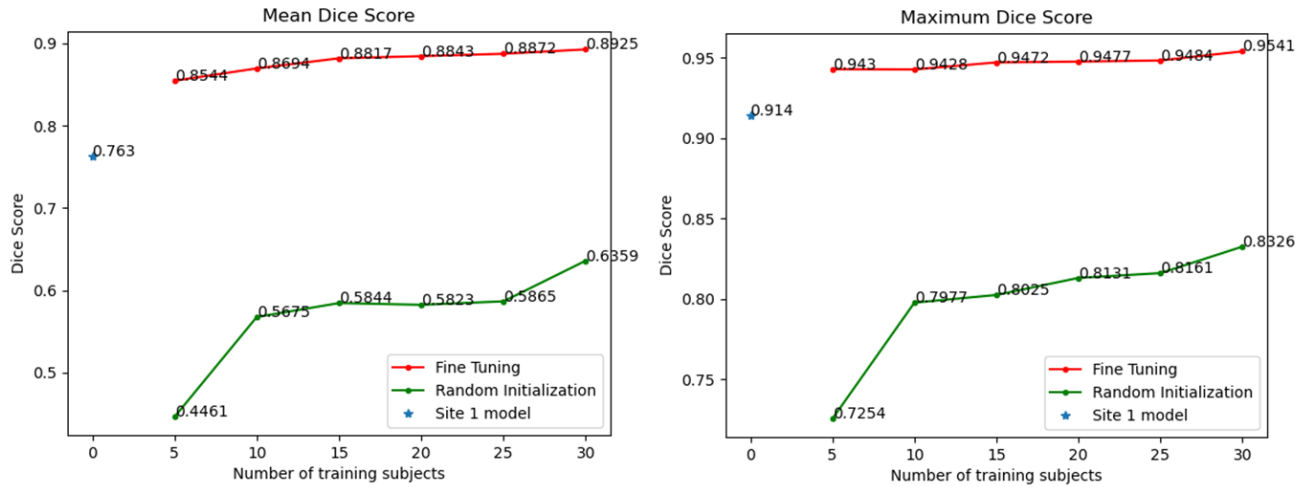

**Figure 4.** The mean and maximum Dice coefficients for two sets of experiments. Plotting of Dice coefficients with random initialization (green) and those when the model is fine-tuned (red) from the previously trained weights demonstrate that as the number of training images increased from 5 to 30, Dice coefficients continue to increase. However, for the fine-tuning case, the increase is less significant after 10 training images.

## 4 Experiments

We conducted two sets of experiments. Each set consisted of six experiments representing consecutive increases in the number of training subjects by 5. During the first set of experiments, the model was initialized using random weights from a normal distribution with mean 0 and standard deviation 1. In the second set of experiments, the model was initialized using TL by pre-training from the initial model (i.e., Site 1).

All training parameters, except the learning rate, were kept constant across all the experiments for a reasonable comparison. The learning rate for the first set was 0.001, whereas the learning rate for the second set of experiments involving fine-tuning was set to 0.0001. The training was continued for 500 epochs with an exponential learning rate decay of 0.9 and using the Adam optimizer for minimizing a multi-class Dice loss function[17]. All the experiments were carried out on an eight-graphics processing unit system (8 V100s on DGX1 system: NVIDIA, Inc., Santa Clara, CA).

Image-mask pairs were produced for each patient using the methods discussed in section 3.2. Out of the 95 cases, 45 were randomly chosen for the test set and 20 for the validation set; the remaining 30 cases were used for training. We conducted six experiments, with systematic varying of the number of cases included in the training set (using 5, 10, 15, 20, 25, and 30 cases). The first group was created by choosing 5 cases randomly from the 30 training cases. In all subsequent groups, 5 randomly chosen cases were added. Thus, each subsequent group was a superset of the previous group. This kind of grouping simulates a continuous addition of new data into the training set.

The performance of the different models was evaluated using the average and maximum Dice scores obtained for the segmentation of LV cavity and LVM. Additionally, Pearson's Correlation Coefficient was utilized to measure the agreement between the ground truth and the predicted values for the LVM wall thickness. A high correlation coefficient (max 1.0) shows a high degree of correlation.

## 5 Results

Increasing the number of training images increased the Dice score for cases where the model weights are randomly initialized (i.e., the model is trained de novo) [Figure 4]. However, in cases where the model is initialized using pre-trained weights by TL, there is no significant gain in the Dice score after increasing the training size beyond 10 subjects.

When the segmentation predictions were made using the original model without any fine-tuning, we found the mean and maximum Dice scores to be 0.763 and 0.914, respectively. As expected, these parameters are higher than the results obtained from models trained on 30 subjects using random initialization (mean 0.6359, max 0.8326) but not as strong as the results obtained by the fine-tuned models (mean 0.8925, max 0.9541) [Figure 4]. Figure 5 shows the improvement in the blood-pool and LVM mask prediction due to transfer learning.

The results of a Pearson's Correlation test on median, 95th percentile and maximum LVM wall thickness measurements in ED and ES across the 45 test subjects are shown [Table 1]. Our results indicate that TL leads to a higher correlation

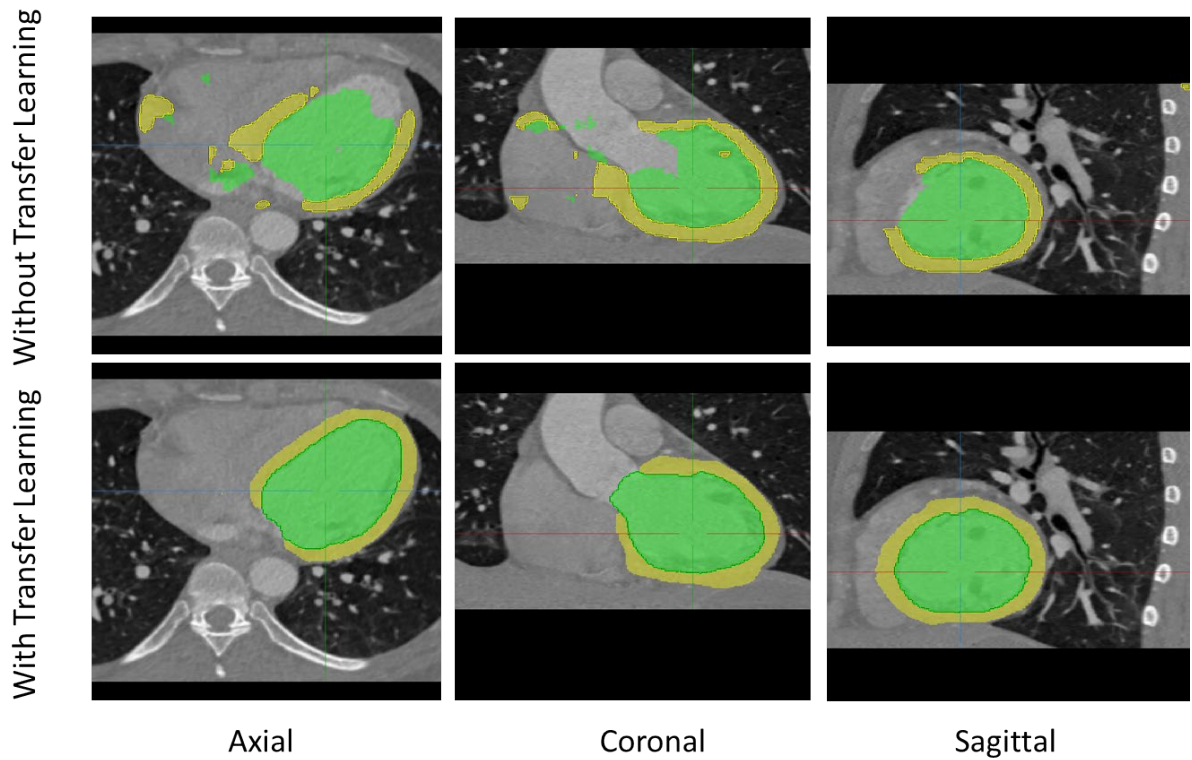

**Figure 5.** Predicted segmentation for LV blood-pool (green) and LVM (yellow). Top row shows that the predictions using a model trained using random initialization [**top row**], are weaker than predictions using a model trained with the benefit of TL [**bottom row**]. Both models are trained using 30 training subjects.

co-efficient when the number of training images is fixed.

## 6 Discussion

In this report, we show that if initiated by TL, an AI model for an imaging use-case can be developed with sparse labels on a small dataset with acceptable levels of accuracy achieved. This process reduces the time required to build a new AI model in the clinical environment at a different institution.

In this report, we discussed some of the challenges in applying such techniques using real-world clinical data. First, there is a legal process involving agreements to which the participating organizations must agree. Second, the guidelines and tools used for annotating images should be compatible for smooth inter-institution operations. In addition, institutions might use different scanners for image acquisition and different DICOM structures for archiving their data; these processes need to be coordinated or appropriately considered.

As discussed earlier, one of the reasons that local engagement in AI is not widespread in Radiology is that many institutions are unable to allocate sufficient resources for labeling image data for training a neural network. We show that, with the help of TL, it is possible to achieve adequate algorithm performance results by labeling a small number of images. Although in this project, the initial model was trained using dense segmentation (i.e., every-other cardiac-cycle phase labeled) at Site 1, only sparse segmentation (only ED and ES phases labeled) was used for fine-tuning at Site 2.

We have shown the effectiveness of the TL process through Dice co-efficient measurement and LVM thickness statistics. Although the Dice co-efficient captures how well the ground truth and the predicted masks overlap, comparing the LVM thickness statistics corroborates our claim that even with a smaller dataset, one can achieve high levels of agreement between the ground truth and the predicted values. We emphasize that fine-tuning of an existing model is necessary, even if the model is being used for the same application; this allows the model to adapt to local variability in data. This possibility is evident in our work from the fact that the Site 1 model did not initially perform well at Site 2 until such fine-tuning. We showed that model fine-tuning with as few as 5 new images significantly boosted the performance of the neural network.

| LVM Wall Thickness (mm) | | |
|---|---|---|
| | Before TL | After TL |
| **End-Diastole** | | |
| Median | 0.72 | 0.74 |
| 95 percentile | 0.64 | 0.69 |
| Max | 0.42 | 0.54 |
| **End Systole** | | |
| Median | 0.72 | 0.92 |
| 95 percentile | 0.81 | 0.90 |
| Max | 0.67 | 0.78 |

**Table 1.** Pearson Correlation Coefficient between the predicted LVM wall thickness measurements and the ground truth before and after transfer learning (TL).

## 7 Conclusion

There are both potential challenges and benefits associated with TL between multiple clinical institutions. This report elucidates both, in preparation for the development of future specific imaging use-cases.